\documentstyle[aps,prl,multicol,amsmath,amstext,amsbsy,amsfonts,amssymb,epsfig]{revtex}

\renewcommand{\v}[1]{{\bf #1}}

\newcommand{\Ref}[1]{Ref.~\cite{#1}}

\newcommand{\Eq}[1]{Eq.~(\ref{#1})}

\newcommand{\vev}[1]{\left\langle #1 \right\rangle}

\renewcommand{\Im}{{\rm Im}}

\newcommand{\prt}{\partial}

\newcommand{\sgn}{{\rm sgn}}
\renewcommand{\t}[1]{{\tilde #1}}

\newcommand{\ie}{{\it ie~}}

\newcommand{\etc}{{\it etc~}}
\newcommand{\etal}{{\it etal~}}

\newcommand{\al}{\alpha}
\newcommand{\bt}{\beta}
\newcommand{\del}{\delta}

\newcommand{\eps}{\epsilon}
\newcommand{\veps}{\varepsilon}

\newcommand{\la}{\lambda}

\renewcommand{\th}{\theta}

\newcommand{\si}{\sigma}

\newcommand{\cL}{ {\cal L} }

\newcommand{\bpm}{\begin{pmatrix}}
\newcommand{\epm}{\end{pmatrix}}
\newcommand{\bmm}{\begin{matrix}}
\newcommand{\emm}{\end{matrix}}

\def\journal #1, #2, #3, 1#4#5#6{{\sl #1~}{\bf #2}, #3 (1#4#5#6) }

\renewcommand{\v}[1]{{\bbox #1}}

\begin{document}
\draft
\widetext

\title{
Continuous topological phase transitions between clean
quantum Hall states
}

\author{Xiao-Gang Wen}
\address{
Department of Physics and Center for Materials Science and Engineering,\\
Massachusetts Institute of Technology,
Cambridge, MA 02139, USA
}

\date{July, 1999}
\maketitle

\widetext
\begin{abstract}
\rightskip 54.8pt
Continuous transitions between states with the {\em same} symmetry but
different topological orders are studied. Clean quantum Hall (QH)
liquids with neutral non-bosonic quasiparticles are shown to
have such transitions under right conditions.
For clean bilayer $(mmn)$ states, a
continuous transition to other QH states (including non-Abelian states)
can be driven by increasing interlayer
repulsion/tunneling. The effective theories describing the critical points
at some transitions are obtained.
\end{abstract}

\pacs{PACS numbers: 73.40.Hm, 73.20.Dx}

\begin{multicols}{2}

\narrowtext


Matters have several different states, such as solid, gas, superfluid, \etc,
under different conditions. According to Landau's theory, all those states
of
matters are characterized by their symmetries.  QH liquids
discovered in 1982\cite{TSG,L} in 2D electron gas form a new state of matter
which contain a completely different kind of order (called topological
order\cite{Wtop}). The topological order is new since it is not
related to symmetries.\cite{Wtop}

One way to gain a better understanding of the states of matters is to study
{\it continuous} transition between different states. For the states
characterized by symmetries, continuous transitions between them
are characterized by a change of symmetry. The transition point
is described by a critical theory which demonstrates various scaling
properties.
Similarly, to gain a better understanding of the
topological orders, we can also study
continuous transitions between different QH liquids.
Since the topological orders are not characterized by symmetries,
some fundamental questions naturally arise:\\
1) Do continuous phase transitions exist
between different QH liquids?\\
2) Are the transition points described by critical theories with scaling
properties? 

In the presence of disorders, the transitions between quantum Hall (QH)
liquids
are believed to be continuous and are described by critical points. But
in this paper, we would like to avoid the complication of disorders and
would
like to concentrate on non-random system.
It was pointed out in \Ref{WW} that a continuous transition between two QH
liquids of difference filling fractions can happen if we turn on a proper
periodic
potential.  This allows us to study continuous phase transitions between QH
liquids without introducing disorders.
The transitions studied in
\Ref{RG,FSW} for paired states are other examples of continuous
transitions (where a periodic potential is not needed).
In this paper we apply \Ref{WW}'s
results to study the transitions between two QH liquids of the same filling
fraction. We will see, in this case, that the transitions some times can be
continuous even without periodic potentials (and disorders).  Such
continuous
transitions between clean QH states can actually appear in bilayer QH
states. We will study some simple bilayer QH states and determine under
which conditions it is likely to observe the continuous phase transition.
The transitions between the paired states\cite{RG,FSW}
are special cases of the transitions studied in this paper.
Since the continuous transitions studied in this paper are
transitions between two HQ states with the {\em same} symmetry, they are
fundamentally different from the usual continuous transitions which change
the
symmetry in the states.  We will call such transitions continuous
topological
phase transitions.


We start with an Abelian QH state described by $(K, \v q)$ with a
Chern-Simons
(CS) effective theory\cite{Kmat}
\begin{equation}
\cL=-\frac{1}{ 4\pi} K_{IJ}a_{I\mu}\prt_\nu a_{J\la}\veps^{\mu\nu\la}
- \frac{e}{ 2\pi} q_I A_{\mu}\prt_\nu a_{I\la}\veps^{\mu\nu\la}
\label{LK}
\end{equation}
where $K$ is a symmetric integer matrix and $\v q$ an integer vector.
The quasiparticles are labeled by integer vectors $\v l$.
The charge and the statistics of such a quasiparticle are given by
\begin{equation}
\th_{\v l}=\pi \v l^TK^{-1}\v l,\ \ \
Q_{\v l}=-e \v l^T K^{-1} \v q
\label{thQ}
\end{equation}

Now assume that we have a gas of quasiparticles labeled by $\v l$ on top
of the $(K,\v q)$ state.
The filling fraction of the quasiparticle gas is
$ \nu_q= \frac{n_q hc}{Q_{\v l} B} $, where $n_q$ is the quasiparticle
density.
If
$\nu_q=\frac{1}{n_e-\frac{\th_{\v l}}{\pi}} $
for an even integer $n_e$, then the quasiparticles can form an
Laughlin state, and we obtain a new Abelian QH liquid labeled by (see Blok
and
Wen in \Ref{Kmat})
\begin{equation}
 K^\prime = \bpm K     & -\v l \\
              -\v l^T & n_e  \epm  ,\ \ \
 \v q^\prime = \bpm \v q\\ 0 \epm
\label{Kqprime}
\end{equation}


Now let us consider a transition induced by condensation
of charge neutral quasiparticles labeled by $\v l$.
First we assume that the quasiparticles labeled by $\v l$ and $-\v l$
(the anti-quasiparticles) have lowest energy gap (so that they control the
transition). The low energy effective theory for the quasiparticles and
the anti-quasiparticles has a form
\begin{eqnarray}
\cL &=&  |(\prt_0 + i a_0 )\phi|^2 -  v^2 |(\prt_i + i a_i )\phi|^2
- m^2 |\phi|^2 - g |\phi|^4
\nonumber\\
&&  -\frac{\pi}{\th_{\v l}} \frac{1}{4\pi} a_\mu\prt_\nu a_\la
\eps^{\mu\nu\la}
\end{eqnarray}
Binding additional $n_e$ flux quanta ($n_e=$even) to the bosonic field
$\phi$
gives us composite boson field $\t \phi$.
The effective theory can also be written
in terms of $\t \phi$
\begin{eqnarray}
\cL &=&  |(\prt_0 + i a_0 + i b_0)\t \phi|^2
-  v^2 |(\prt_i + i a_i +i b_i)\t \phi|^2
- m^2 |\t \phi|^2
\nonumber\\
&&
- g |\t \phi|^4
-\frac{\pi}{\th_{\v l}} \frac{1}{4\pi} a_\mu\prt_\nu a_\la \eps^{\mu\nu\la}
+\frac{1}{n_e} \frac{1}{4\pi} b_\mu\prt_\nu b_\la \eps^{\mu\nu\la}
\label{Leff}
\end{eqnarray}
Here $2m$ is the energy gap for creating a quasiparticle-anti-quasiparticle
pair.

\begin{figure}
\epsfysize=1.2truein
\centerline{ \epsffile{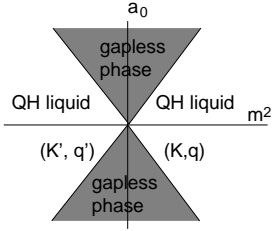} }
\caption{
The phase diagram near a critical point.
}
\label{fig1}
\end{figure}

Near the transition, two parameters $m^2$ and $a_0$ are important.
(We can always assume $b_0=0$ without losing generality.)
The (mean field) phase diagram is sketched in Fig. \ref{fig1}.
The transition at $a_0=m^2=0$ is the transition from
the $\vev{\t \phi}=0$ phase (the $(K,\v q)$ phase) to the
$\vev{\t \phi}\neq 0$ phase
(the $(K^\prime,\v q^\prime)$ phase).
Both $(K,\v q)$ and
$(K^\prime,\v q^\prime)$ phase have finite energy gap, while gapless neutral
excitations appear at the transition point.
The effective theory \Eq{Leff} for the transition is identical to the one
studied in \Ref{WW}.
It was shown that the transition is continuous at least
in a large $N$ limit.\cite{WW}
The transition point is a critical point with scaling properties.
Some critical exponents at the transition point
($a_0=m^2=0$) were calculated in the large $N$ limit.

%

We see that a continuous
transition between the two QH liquids $(K,\v q)$ and $(K^\prime,\v
q^\prime)$
can happen even without the lattice if $Q_{\v l}=0$.
{\em a QH state with neutral
quasiparticles can have a continuous phase transition to another QH state
even in the clean limit.} One can show that the two QH states always have
the
same filling fraction.

We can also bind an odd number $n_o$ flux quanta to $\phi$ and use an
effective composite fermion theory to describe the transition
\begin{eqnarray}
\cL &=&
\psi^\dag (\prt_\mu + i a_\mu +ib_\mu) \tau^\mu \psi
- m \psi^\dag \tau^3 \psi
\nonumber\\
&& +\Psi^\dag (\prt_\mu + i a_\mu +ib_\mu) \tau^\mu \Psi
- M \Psi^\dag \tau^3 \Psi
\nonumber\\
&&  -\frac{\pi}{\th_{\v l}} \frac{1}{4\pi} a_\mu\prt_\nu a_\la
\eps^{\mu\nu\la}
+\frac{1}{n_o} \frac{1}{4\pi} b_\mu\prt_\nu b_\la \eps^{\mu\nu\la}
\label{LeffF}
\end{eqnarray}
where $\tau^0=1$ and $\tau^i|_{i=1,2,3}$ are the Pauli matrices and $\psi$
and
$\Psi$ are two-component fermion fields. $M$ is a large number and
$\Psi$ is the regularization field.
Integrating out $\psi$ and $\Psi$ generates a CS term
$ -\frac{\sgn(M)\th(mM)}{4\pi} a_\mu\prt_\nu a_\la \eps^{\mu\nu\la} $.
The (mean field) phase diagram for composite fermion effective theory
is identical to that of composite boson (Fig. \ref{fig1}),
if we replace $m^2$ by $-mM$.
The critical properties of \Eq{LeffF} at $a_0=m=0$ are studied in \Ref{CFW}.
The transition was shown to be continuous in the limit $\th_{\v l}\to \pi$
and $n_o=0$.
Both \Eq{Leff} and \Eq{LeffF} describe the same set of transitions, labeled
by
$n_e$ or $n_o$.
The two sets of labels are related by $n_e=n_o-\sgn(M)$.

In general, after transition, the new QH liquid described by
$(K^\prime,\v q^\prime)$ contains
dim$(K)+1$ edge branches, since dim$(K^\prime)={\rm dim}(K)+1$.
The quantum number of the quasiparticles are given by \Eq{thQ}.
However,
when $(K^\prime,\v
q^\prime)$ contains neutral null vectors,\cite{H,KCW}
the  $(K^\prime,\v q^\prime)$ QH liquid is really a QH liquid described by a
reduced  $(\t K,\t{\v q})$ with dimension dim$(K)-1$.\cite{H,KCW}
A calculation of  $(\t K,\t{\v q})$ from $(K^\prime,\v
q^\prime)$ was out lined in \Ref{H}.
We would like to point out that if
$n_e=0$ in \Eq{Kqprime}, $(K^\prime,\v q^\prime)$ always has at least one
neutral null vector $\v l_{null}^T= (\v l^T, 0 )$, and
the transition reduces the number of edge branches.


Now let us study a concrete bilayer $(nn0)$ state
to gain more detailed understanding of the transitions.
Here $n$ is odd if electrons are fermions, and $n$ is
even if electrons are bosons.
The electrons form a $\nu=1/n$ Laughlin state in each layer.
The quasiparticle labeled by $\v l^T= ( -1,1)$ is neutral and has
statistics $\th_{\v l}= 2\pi/n$.
Such a neutral quasiparticle corresponds to a bound
state of the charge $-e/n$ Laughlin quasiparticle in one layer and  the
charge
$e/n$ Laughlin quasihole in the other layer.
If the Laughlin quasiparticle and quasihole in the two layers have strong
enough attraction (this happens when there is a strong interlayer
repulsion),
the quasiparticle-quasihole pair (\ie the
neutral quasiparticle labeled by $\v l^T= (-1,1)$ and the neutral
anti-quasiparticle  labeled by $\v l^T= (1,-1)$) will be spontaneously
generated.  This will cause a phase transition.

To understand the nature of the transition, let us first assume that the
neutral (anti-)quasiparticles are bosons with $\th=0$. The dynamical
properties of the neutral (anti-)quasiparticles can be modeled by a lattice
boson system which only allow 0, 1, or 2 bosons per site. The average boson
density is one boson per site. In the lattice-boson model, an empty site
corresponds to the neutral anti-quasiparticle, a double occupied site
corresponds to the neutral quasiparticle, and a singly occupied site
correspond to no quasiparticle. Let us first ignore the boson hopping. When
the interlayer repulsion is much weaker than the intralayer repulsion, the
ground state of the lattice-boson model has exactly one boson per site,
and is a Mott insulator.  If the interlayer repulsion is much
stronger than the intralayer repulsion, the lattice-boson model will has two
degenerate ground states, one has two bosons per site and the other has no
boson.  This corresponds to a charge unbalanced state where electrons have
different densities in the two layers.  Such a charge unbalanced state has
been observed in experiments.\cite{ch-asymm}
In the presence of boson hopping, before going into the charge unbalanced
phase, the boson Mott insulator must undergo a continuous phase transition
into a boson superfluid phase. (The effective theory near the boson
Mott-insulator to  superfluid transition is given by \Eq{Leff} without the
gauge fields.)
This is the phase transition between the $K$
and $K^\prime$ states.
In the boson condensed phase, the bosons condense into a state described by
a
constant wave function $\Phi(\{z_i\}; \{w_i\})$=Const., where the
complex number $z_i$ ($w_i$) are coordinates of the neutral quasiparticles
(anti-quasiparticles).

When $\th\neq 0$, the phase diagram is
similar to the one for $\th=0$ (at least in the large $N$ limit\cite{WW}),
and the picture discussed above still applies.
The effective theory near the transition is given by \Eq{Leff} with the CS
term.
In general the (anti-)quasiparticle can condense into a
state described by wave function
\begin{equation}
\prod_{i<j} \left[(z_i-z_j)(w_i-w_j)\right]^{n_e-\frac{\th}{\pi}}
\prod_{i,j} (z_i^*-w_j^*)^{n_e-\frac{\th}{\pi}},
\end{equation}
where $n_e$ is a
certain even integer which corresponds to the $n_e$ in \Eq{Leff}.

In the dilute limit, the state with minimal $\left| n_e-\frac{\th}{\pi}
\right|$ is expected to has lowest energy.
Therefore, if $n\ge 2$,
as we increase the interlayer repulsion, the
$(nn0)$ state will transform into a QH state with
\begin{equation}
K_3 = \bpm n&0&1\\ 0&n&-1\\1&-1&0 \epm ,\ \ \ \v q_3 = \bpm 1\\1\\0 \epm
\label{K3q3}
\end{equation}
via a continuous transition.

What is the $(K_3, \v q_3)$ state? It is nothing but the
$(K,\v q)=(2n, 2)$ state,
\ie the $\nu=1/2n$ Laughlin state of charge-$2e$ bosons
(the Hall conductance is
$\si_{xy}=\nu (2e)^2/h$). This is because after a $SL(3,Z)$ transformation,
the $(K,\v q)$ in \Eq{K3q3} can be rewritten as
\begin{equation}
K = \bpm 2n&0&0\\ 0&n\% 2 &-1\\0&-1&0 \epm ,\ \v q = \bpm 2\\1\\0 \epm
\label{K3q3a}
\end{equation}
where $ n\% 2=1$ if $n=$odd and $ n\% 2=0$ if $n=$even.
According to Haldane's topological instability\cite{H,KCW},
\Eq{K3q3a} just describes the $(K,\v q)=(2n, 2)$ state.
On the edge, the three edge branches described by $K_3$ in \Eq{K3q3a}
can be viewed as the reconstructed edge of
the  $(K,\v q)=(2n, 2)$ state.\cite{KCW}

When $n=2$, the effective theory \Eq{Leff} can be mapped into a free fermion
model (since the interaction terms are irrelevant)
\begin{eqnarray}
\cL &=&
\psi^\dag \prt_\mu \tau^\mu \psi
- m \psi^\dag \tau^3 \psi
\label{LeffF2}
\end{eqnarray}
The effective free fermion theory allows us to calculate all the physical
properties of the transition, and we can show rigorously the transition to
be
continuous.

The effective theory \Eq{Leff} used to describe the transition
has a $U(1)$ symmetry -- the conservation of the neutral
quasiparticle $\phi$.
However, such a $U(1)$ symmetry is broken by interlayer electron tunneling
which create $n$ neutral (anti)quasiparticles.
The electron tunneling operator has a form $\hat M\phi^n$, where
$\hat M$ is an operator that create one unit of $a_\mu$
flux. Note that the combination $\hat M\phi^n$ is gauge invariant and
the effective Hamiltonian/Lagrangian \Eq{Leff} should contain a term
$t \hat M\phi^n+h.c.$ to describe the electron interlayer tunneling.

When $n$ is large, we expect the tunneling term $t \hat M\phi^n+h.c.$
to be irrelevant and it can affect the properties of the transition only if
$t$
is large.
When $n$ is small the effect of the tunneling term
may be important even in small
$t$ limit. When the tunneling term is important, its effect is hard
to study. However, since  the effective theory for $n=2$
can be mapped into a free fermion model, the
effect of the tunneling term on the transition can be studied exactly.
This situation has been studied in \Ref{RG,FSW} for a related $(331)$ state
(or more general $(q+1,q+1,q-1)$ state in \Ref{RG}).
In the following we will show how to make contact with their derivation.

\begin{figure}
\epsfxsize=3.2truein
\centerline{ \epsffile{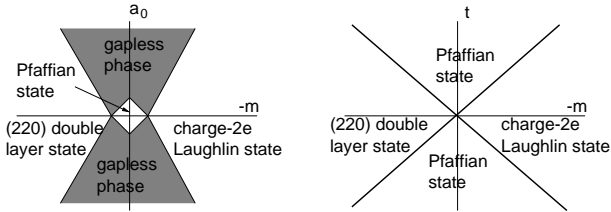} }
\caption{
The phase diagram for the $(220)$ state in the presence of
interlayer tunneling. We have assumed $\Im t=0$.
The left diagram has a fixed $t$ and the right one has $a_0=0$.
}
\label{fig3}
\end{figure}

For $n=2$, we start
with the fermionic effective theory
\Eq{LeffF2}. Since the electron tunneling operator
create a pair of quasiparticles $\psi$,
it has a form $ \psi^T \tau^2 \psi$. Thus the effective
Lagrangian with tunneling can be written as
\begin{eqnarray}
\cL &=&
\psi^\dag \prt_\mu \tau^\mu \psi
- m \psi^\dag \tau^3 \psi
+ (t \psi^T \tau^2 \psi + h.c.)
\label{LeffF2g}
\end{eqnarray}
where $t$ is the amplitude of the interlayer electron tunneling.
After diagonalization, the Hamiltonian becomes
$ H=\sum_{\v k, \al=\pm} E_\al (\v k) \la_{\al, \v k}^\dag \la_{\al,\v k} $
with
\begin{equation}
 E_\pm(\v k)= \sqrt{ k^2 +m^2+|t|^2 \pm 2 \sqrt{m^2|t|^2 +k^2 (\Im t)^2}}
\end{equation}
The system contains gapless excitations (\ie reaches a critical point)
when $m=|t|$ or $m=-|t|$. We see that the single transition point is split
into
two transition points by the tunneling term. The new phase diagram is
sketched in
Fig. \ref{fig3}.  Near the new transition points the
low energy excitations are described by one free gapless Majorana
fermion
$H= \sum_{\v k} \sqrt{ v^2 \v k^2 + (|t|-|m|)^2} \la_{\v k}^\dag \la_{\v
k}$,
where $v = 1- \frac{ (\Im t)^2 }{|m t|}$.
This agrees with the result obtained in \Ref{RG,FSW}.
Again, all the physical properties of the transition can be calculated from
the above free fermion effective theory.
The state between the two new transition points\cite{RG,FSW} is a
non-Abelian Pfaffian state (for bosons) proposed by
Moore and Read.\cite{MR}
It was suggested\cite{GWW} that such a $p$-wave paired state may describe
the
$\nu=5/2$ state observed in experiments.

{}From the phase diagram Fig. \ref{fig3} (which has been given in \Ref{RG}),
we see that the
continuous transition from the $(220)$ state to the non-Abelian
Pfaffian state can also be induced by increasing the interlayer tunneling
$t$. One naturally asks
what kind of QH state can a large $t$ induce from the $(nn0)$
state?

We first note that the analytic part of the $(nn0)$ ground state wave
function
can be written as a correlation function of 2+0D chiral
fermions.\cite{MR,WWH}
The chiral fermions $\psi_i$ is defined by operator product expansion (OPE)
$\psi_i^\dag(z)\psi_j(0)=\del_{ij}/z$ where $z=x+iy$. Introduce
electron operators in the two layers
$\psi_{e1}=\prod_{i=1}^n\psi_i$, and
$\psi_{e2}=\prod_{i=n+1}^{2n}\psi_i$, the $(nn0)$ wave function can be
written
as
$\Phi_{(nn0)}(\{z_i\};\{ w_i\})=\vev{e^{-iN\phi} \prod_{i=1}^N
\psi_{e1}(z_i)\psi_{e2}(w_i)}$,
where the boson field $\phi$ is defined by
$\frac{1}{2\pi}\prt_z\phi(z)=\sum_{i=1}^{2n}\psi_i^\dag(z)\psi_i(z)$.

For the $(220)$ state, after the transition, the wave function of the
resulting Pfaffian state is just the $(220)$ wave function symmetrized
between
$z_i$ and $w_j$. Such a wave function can be written as a correlation
function
of a single electron operator $\psi_e=\psi_{e1}+\psi_{e2}$:
\begin{eqnarray}
&& \Phi_{\rm nab}(z_1,...,z_{2N})=\vev{e^{-iN\phi} \prod_{i=1}^{2N}
\psi_{e}(z_i)}
\label{Phi_nab}
\end{eqnarray}
Since only a single electron operator is involved,
$\Phi_{\rm nab}$ is a single layer state. This is consistent with the fact
that
the $\Phi_{\rm nab}$ state is induced by a large interlayer tunneling.

Here we would like to conjecture that the similar phenomenon will also
happen
for the $(nn0)$ state with $n>2$.
A large interlayer tunneling $t$ will
change the $(nn0)$ state into the $\Phi_{\rm nab}$
state defined in \Eq{Phi_nab}.

When $n=3$, we find that the $\Phi_{\rm nab}$ state is a non-Abelian state.
The edge excitations are generated by $\psi_e$'s and
$\psi_e^\dag$'s.\cite{WWH}
Through the OPE of  $\psi_e$ and $\psi_e^\dag$, one can show that
the neutral edge excitations are generated by $\rho_{+}$, $\rho_{-}^2$
and $\cos(3n\phi_-)$,
where $\rho_\pm$ are the currents in $U(1)^2$ Kac-Moody algebra and
$\phi_\pm$ are defined by $\prt_z \phi_\pm/2\pi=\rho_\pm$. Physically
$\rho_+$ corresponds to the total electron density and $\rho_-$ corresponds
to
the difference of the electron densities in the two layers.
They have the following OPE $\rho_\al(z)\rho_\bt(0)=2\del_{\al\bt}/3z^2$.
The electron operator can be written as
$\psi_e=e^{i3\phi_+/2} \cos(3\phi_-/2)$ through bosonization. Thus the
electron propagator alone the edge has an exponent 3:
$\vev{\psi_e\psi_e^\dag}\sim z^{-3}$. The quasiparticle operators must be
mutually local with respect to
the electron operator.\cite{Wtop} We find the quasiparticles with
lowest charge is created by $\psi_q=e^{i\phi_+/2} \cos(\phi_-/2)$ which
carries charge $e/3$. The quasiparticle propagator has an exponent 1/3.
The electron and the quasiparticle exponents and charges (and hence the edge
tunneling I-V curve) for our single layer $\nu=2/3$ non-Abelian state
are identical with the $\nu=1/3$ Laughlin state.

Since all the charged excitations remain to have finite energy gap across
the
transition, the continuous topological phase transitions discussed in this
paper may not be easy to observe.  Notice that the neutral gapless
excitations at the transition carry an electric dipole moment in the
$z$-direction.
Thus one way detect them is to use surface acoustic phonon.
Also the edge states before and
after the transition are very different. Near the transition, the velocity
of one edge mode approaches to zero, and such mode becomes the gapless bulk
excitations at the transition. Thus the transition should also be detectable
through edge tunneling experiments.

We would like to mention that
the phase diagram for the $(n,n,m)$ state is the same
as that of the $(n-m,n-m,0)$ state discussed above. The neutral
excitations in the two states are identical. As a consequence, the critical
theories for the transitions in the two states are identical.
The results in this paper can be easily generalized
to any $(mnl)$ bilayer states and hierarchical states.

XGW is supported by NSF Grant No. DMR--97--14198 and by NSF-MRSEC Grant
No. DMR--98--08941.

\end{multicols}

\end{document}